# Sending Hidden Data via Google Suggest

Piotr Białczak, Wojciech Mazurczyk, Krzysztof Szczypiorski
Warsaw University of Technology, Faculty of Electronics and Information Technology, Institute of Telecommunications, 15/19 Nowowiejska Str.
00-665 Warsaw, Poland
E-mail: P.Bialczak@stud.elka.pw.edu.pl,
{W.Mazurczyk,K.Szczypiorski}@tele.pw.edu.pl

*Abstract*

Google Suggest is a service incorporated within Google Web Search which was created to help user find the right search phrase by proposing the auto-completing popular phrases while typing. The paper presents a new network steganography method called StegSuggest which utilizes suggestions generated by Google Suggest as a hidden data carrier. The detailed description of the method's idea is backed up with the analysis of the network traffic generated by the Google Suggest to prove its feasibility. The traffic analysis was also performed to discover the occurrence of two TCP options: Window Scale and Timestamp which StegSuggest uses to operate. Estimation of method steganographic bandwidth proves that it is possible to insert 100 bits of steganogram into every suggestions list sent by Google Suggest service.

## 1 Introduction

The main aim of network steganography is to hide secret data (steganograms) in the normal transmissions of users. In ideal situation hidden data exchange cannot be detected by third parties. Network steganography may be easily utilized as a tool for data exfiltration or to enable network attacks [2].

Contrary to typical steganographic methods that utilize digital media (pictures, audio and video files) as a cover for hidden data (steganogram), network steganography utilizes communication protocols' control elements and their basic intrinsic functionality. Typical steganographic methods have proven to be useful tools for data exfiltration e.g. in 2008 IEEE Spectrum magazine [3] wrote that "someone at the United States Department of Justice smuggled sensitive financial data out of the agency by embedding the data in several image files"; this year the same magazine [4] reported that the Russian spy ring was uncovered which used similar steganographic techniques. However embedding secret data into images has two serious drawbacks: it allows hiding limited amount of data per one file and the modified picture may be accessible for forensics experts (e.g. because it was uploaded to some kind of server). With network steganography it is different; it allows leaking information (even very slowly) during long periods and if all the exchanged traffic is not captured than there is nothing left for forensics experts to analyse. As a result, such methods are harder to detect and eliminate from networks. In order to minimize the potential threat to public security, identification of such methods is important as the development of effective detection (steganalysis) methods. This requires both in-depth understanding of the functionality of network protocols and ways in which it can be used for steganography.

The best carrier for secret messages (steganograms) must have two features. Firstly, it should be popular i.e. usage such carrier should not be considered as an anomaly itself. The more such carriers are present and utilized in network it's the better because they mask using the carrier to perform hidden communication. Secondly, modification of the carrier related to inserting the steganogram should not be "visible" to third party not aware of the steganographic procedure.

And how to find a carrier in the network traffic that would fill above-mentioned requirements? In the Internet today we witness expansion of different, advanced Internet services from e-commerce such as Amazon, eBay to information and social web sites (like Twitter, Flickr, MySpace, iGoogle, Wikipedia, Facebook). They all incorporate advanced Web 2.0 mechanisms [1] for customizable content presentation, sharing and delivery and they all have tremendous number of users. And they all use network protocols, sometimes very complex, to realize these services. Thus, they are perfect candidates for secret data carriers.

In this paper we propose a new steganographic method StegSuggest, which exploits popular Google Suggest service traffic to perform hidden communication.

The rest of the paper is structured as follows. Section 2 introduces Ajax (Asynchronous JavaScript and XML) group of technologies which was used to create Google Suggest service. Section 3 describes related work. Section 4 presents experimental results for real-life LAN (Local Area Network) traffic which permit for an evaluation of feasibility of the proposed solution. Section 5 discusses the proposed information hiding system in detail. Finally, Section 6 concludes our work.

## 2 Ajax and Google Suggest Service

Ajax is a group of interrelated web development techniques used on the client-side to create interactive web applications. Ajax incorporates technologies like: HTML/XHTML and CSS (Cascading Style Sheets), Document Object Model and XML (Extensible Markup Language), XMLHttpRequest object for asynchronous communication and JavaScript. Web applications using Ajax are able to retrieve data from the server asynchronously without disrupting the display and behaviour of the currently loaded page.

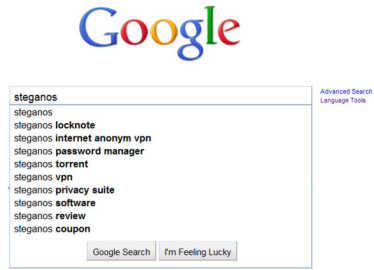

**Fig. 1** Google Suggest example for search phrase *steganos*

Google Suggest is a service created using Ajax and incorporated within Google Web Search – the most popular web search engine in the Internet.

**Fig. 2** Google Suggest HTTP GET request message

Google Suggest was created to help user find the right search phrase by proposing auto-completing popular phrases while typing (Fig. 1). From the network protocol perspective Google Suggest uses HTTP and TCP protocols, the same as are used for typical web page exchange. The algorithm behind this service works as follows:

When user begins typing a search phrase into the search field, once in a while, these characters are sent to Google server in HTTP GET message (Fig. 2).

Google server returns a list of the most popular search phrases in HTTP OK message (Fig. 3). The list is then presented to the user in a drop-down list.

**Fig. 3** Google Suggest HTTP OK message with proposed suggestions

While the user is typing, HTTP GET messages are sent frequently (Fig. 4). This means that during search for particular phrase numerous of HTTP messages will be exchanged (GETs with user typed characters and OKs with suggestions in return).

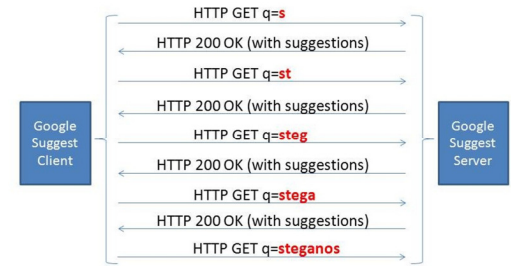

**Fig. 4** Exchange of HTTP GET and OK messages during typing the search phrase *steganos*

Each HTTP request may carry at least one or more characters of the search phrase depending on the user's typing speed (Fig. 5).

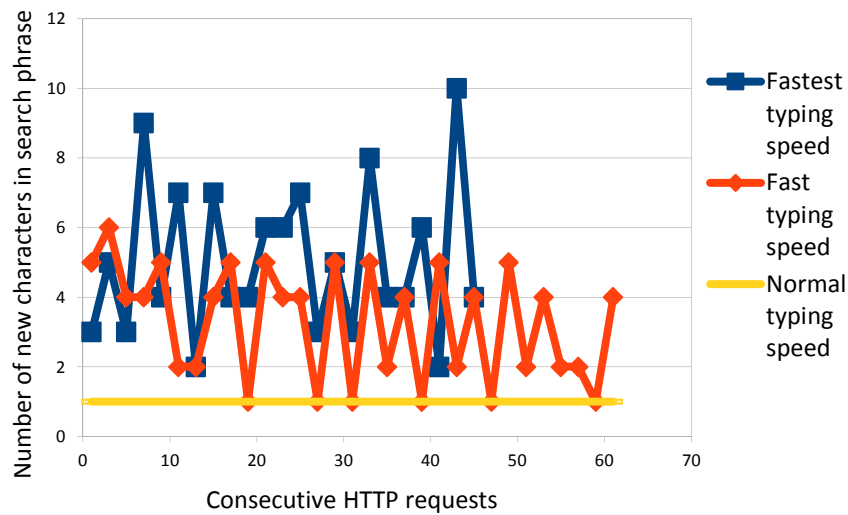

Fig. 5 The number of the new characters in HTTP requests depending on user's typing speed

## 3. Related Work

To authors' best knowledge there are no steganographic methods proposed for such services like Google Suggest. However, proposed steganographic method, StegSuggest, modifies mainly two network protocols: con-tent of HTTP messages containing suggestions and TCP header options: Windows Scale (WS) and Timestamp (TS), and for both protocols steganographic methods were proposed. This section review steganographic method that may be applied to TCP header and HTTP protocol messages.

**TCP**
Giffin et al. in [5] proposed steganographic method which utilizes TCP timestamp header option, which is commonly used to improve TCP performance. The steganograms are inserted into the low order bits of the sender timestamps (similar method is a part of StegSuggest). In [6] Handel and Sanford proposed to enable secret communication by using the unused bits of the TCP header's Flags field. Hintz proposed a method that transmits steganograms in the TCP Urgent Pointer [7]. In [8] Fisk et al. introduced steganographic methods which modify TCP Reset segments and TCP header padding. For TCP ISN (Initial Sequence Number) few steganographic methods were proposed e.g. by Rowland [9], Rutkowska [10] and Murdoch et al. [11]. Abad de-scribed on the example of IP header how to utilize Checksum field for covert communication [12] – the same method is applicable also for the TCP header checksum.

**HTTP**
In [13] Bauer proposed hiding information in JavaScript/HTML and transporting it through the use of JavaScript redirects. Various methods for utilizing HTTP protocol headers for steganographic purposes were proposed by Dyatlov et al., Kwecka and Van Horenbeeck [16–18]. In [19] Castro et al. introduced new steganographic method which uses HTTP cookies to hide steganograms. Feamster et al. in [14] proposed sending covert requests for web pages encoded as a sequence of HTTP requests to innocent web sites and return the content hidden inside harmless images. Bowyer in [15] proposed a similar technique to enable communication with Trojans behind firewalls.

## 4. Google Suggest in Real Network Traffic

To analyse the characteristic and the volume of traffic that Google Suggest generate the real network traffic was captured from LAN. The experiment was conducted at the Institute of Telecommunications at Warsaw University of Technology between 5 of November and 14 of December 2010 (from Mondays to Fridays). The traffic was captured with the aid of Dumpcap which is part of the Wireshark sniffer ver. 1.3.3 (www.wireshark.org). The sources of traffic were ordinary computer devices placed in several university laboratories and employees' ones but also peripherals, servers and network equipment. To analyse the captured traffic and calculate statistics TShark (which is also part of Wireshark) was utilized.

**4.1 HTTP requests**
We want to explore how many Google searches i.e. looking for the particular search phrases, are performed for average user, how many Google Suggest HTTP requests are involved during average search and how many searches per hour does average user generate?

The analysis of the real Google search requests revealed that there were nine types of request which depended on where Google client resided: *hp* (the search was initiated on main Google web search page), *serp* (the search was performed on previously returned search results), *tbrs* (the search was initiated from installed Google toolbar), *firefox*, *ie*, chrome, safari (the search was performed using web browsers and embedded Google toolbar), *img* (the search target were digital images), *youtube* (the searches come from youtube.com site). The total number of HTTP requests and for each Google client is presented in Table 1.

**Table 1** Google Suggest HTTP requests statistics in captured traffic

| Client | Total | hp | serp | firefox | safari | chrome | youtube | ie | tbrs | img | unspec. |
|---|---|---|---|---|---|---|---|---|---|---|---|
| No. of request | 18921 | 7857 | 4830 | 1616 | 1146 | 1057 | 985 | 945 | 354 | 95 | 36 |
| [%] | 100 | 41.53 | 25.53 | 8.54 | 6.06 | 5.59 | 5.21 | 4.99 | 1.87 | 0.50 | 0.19 |

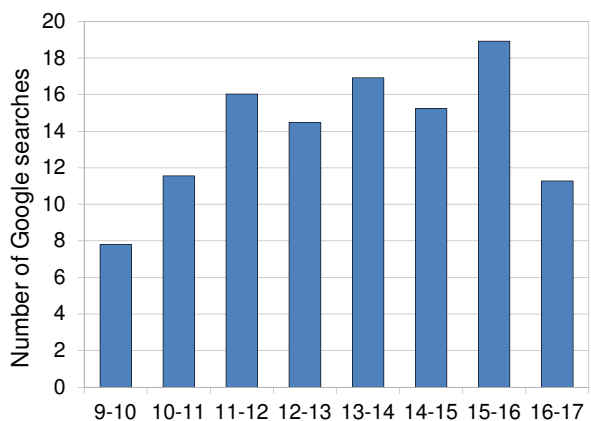

**Fig. 6** The number of average Google searches for different time of day (from 9 am to 5 pm)

From captured LAN traffic we took into account only traffic from users which generated less than 5 Google searches. For the remaining users the following results were acquired which are presented in Table 2.

For calculating number of Google searches per hour we chose only working hours between 9 am and 5 pm. Results reveal that average user generates about 61 searches, where each consists, on average, of about 7 HTTP messages. User generated single Google search, on average, every three hours, which is in fact quite surprising as it is such popular tool. The average Google searches activity of users during different time of the day is presented in Fig. 6.

**Table 2** Google Suggest statistics in captured traffic (averages)

|  | No. of Google searches/user | No. of HTTP requests/user | No. of HTTP req./Google search | No. of Google searches/h |
|---|---|---|---|---|
| No. | 61.4 | 415 | 7.1 | 0.31 |
| Stand. Dev. | 107.1 | 745 | 3.5 | 0.53 |

### 4.2 TCP *Windows Scale* and *Timestamp* options

Traffic analysis was also performed to establish occurrence of the TCP segments related to Google Suggest service (with SYN or SYN with ACK flags) that had options *Window Scale* (WS) and *Timestamp* (TS) set.

The *Window Scale* option can be attached only to a SYN segment. It has two purposes [**Error! Reference source not found.**]:
- indicates that the TCP is prepared to perform both send and receive window scaling,
- advertises a scale factor to be applied to its receiving window; TCP segment that is prepared to scale windows should send the option, even if its own scale factor is 1 (the scale factor is limited to a power of two and encoded logarithmically).

The TS option is used for two mechanisms: RTTM (Round Trip Time Measurement) and PAWS (Protect Against Wrapped Sequences) [21].

Detailed results of the analysis are presented in Tables 3 and 4.

**Table 3** TCP WS and TS options occurrence

|  | SYN | SYN with WS set | SYN with TS option |
|---|---|---|---|
| No. | 3989 | 2422 | 1032 |
| [%] | 100 | 60.7 | 25.9 |

**Table 4** TCP WS values and TS presence

| WS values | SYN with WS value [%] | SYN with WS val. and TS option [%] |
|---|---|---|
| 0 | 0 | 0 |
| 1 | 14.46 | 86.31 |
| 2 | 30.38 | 1.57 |
| 3 | 2.11 | 100 |
| 6 | 10.29 | 100 |
| 7 | 0.53 | 100 |
| 8 | 2.96 | 0 |

It turned out that from all TCP SYN segments around 60% had WS option and around 26% of segments had TS option set. Each segment with TS option set had also WS option. For SYN segments with WS values equal 3, 6 and 7 all had TS option active, for WS 0 and 8 none.

## 5. StegSuggest: Communication Scenarios and Idea

To enable hidden communication StegSuggest utilizes traffic generated by Google Suggest. Its main innovation is to insert new words into suggestions sent to the Google Suggest client.

Inserted words carry bits of steganogram. In this section we describe hidden communication scenarios for StegSuggest (5.1), how steganogram is en(de)coded into original Google suggestions (5.2), how StegSuggest operates in detail (5.3), and finally we estimate its potential bandwidth (5.4).

### 5.1 Steganographic communication scenarios

All possible hidden communication scenarios for StegSuggest are presented in Fig. 7. From these four scenarios only (a) and (b) are feasible ones. In both scenarios some intermediate network node is a steganogram sender and steganogram receiver is also intermediate network node (a) or Google Suggest client (b).

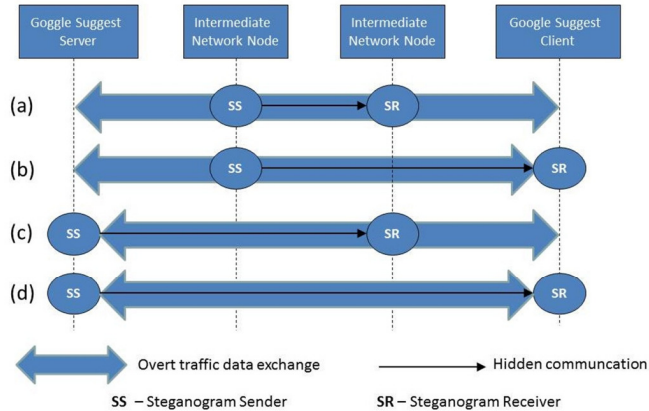

**Fig. 7** Possible hidden data transmission scenarios

It is worth noting that in scenario (a) both sides of overt communication (Google Suggest server and client) are not aware of hidden communication.

Moreover, in this scenario it is possible that steganograms sender (SS) and steganogram receiver (SR) can utilize the whole Google Suggest traffic coming from particular LAN for steganographic purposes and thus achieve higher steganographic bandwidth.

In scenarios (c) and (d) it is assumed that Google Suggest server takes part in steganographic communication as a steganogram sender. Unless such server is a victim of some kind of the network attack such situation is unlikely to happen.

In the rest of the paper StegSuggest is described for scenario (a) from Fig. 7. It is assumed that SS and SR are capable of capturing the whole Google Suggest traffic between Google server and clients that are located e.g. in particular LAN. This gives opportunity for hidden communication based on aggregated Google Suggest traffic which originates from many Google Suggest clients and is destined for one or more Google servers (Fig. 8).

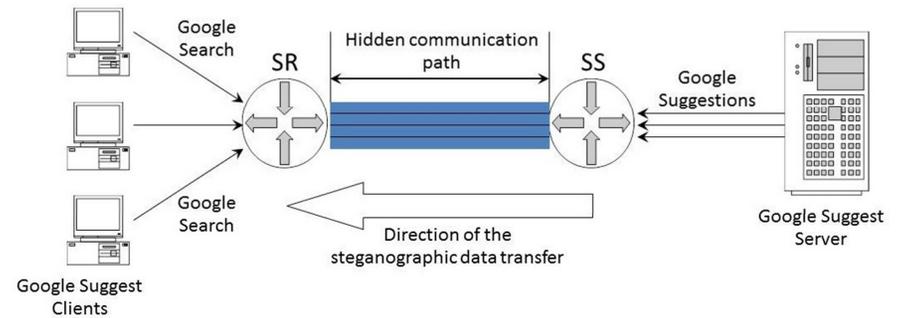

**Fig. 8** StegSuggest transmission scenario

### 5.2 Steganogram encoding and decoding

StegSuggest utilises original suggestions sent by the Google Suggest server to transfer steganograms. It is achieved by adding words to the original suggestions sent to the Google client. Adding new words to the generated suggestions by Google Suggest service is hard to detect because sometimes the suggestions are strange and/or unexpected (examples – Fig. 9 – many Internet sites are devoted exclusively to track such meaningless suggestions).

**Fig. 9** Examples of the strange and funny Google suggestions (http://www.boredpanda.com)

The key issue for StegSuggest to be able to evade disclosure is to design a proper codebook from which the steganographic words (which we called steg-suggestions) will be chosen. Such codebook should fulfil following two requirements:

- R1: only the most popular word in given language will be utilised and it should be hard to distinguish whether these words are added with steganographic purpose or they are strange Google Suggest service suggestions (like e.g. in Fig. 9),
- R2: the same sequence of the bits of steganogram will not always be encoded with the same steg-suggestion word.

To achieve these goals the following steps were taken while designing StegSuggest codebook (Fig. 10). Site *www.wordfrequency.info* contains a *Frequency Dictionary of Contemporary American English* providing a list of the 5000 most frequently used words in the language [20]. The dictionary is based about 400 million words corpus evenly balanced between spoken English (from radio and TV shows), fiction (books, short stories, movies scripts), more than 100 popular magazines, 10 newspapers and 100 academic journals). The idea of the hidden data transfer using StegSuggest is to add words from the abovementioned top 5000 most frequently used English words to the original suggestions sent by Google. Each added word would carry some bits of steganogram. This would ensure fulfilling requirement R1.

First, from the total of 5000 words we selected 4757 words (2542 nouns, 1001 verbs, 839 adjectives, 340 adverbs and 35 quantifiers). The rest of the words like pronouns and prepositions were omitted because using them can be considered as suspicious because the steg-suggestions are added as the last word at each Google suggestion row sent. Second, the homographs (words that share the same written form but have different meanings) were deleted which left 4220 words. Next, we limited the list to 4096 most frequently used words. Based on these results the codebook was created, which consisted of four groups 1024 words each. The words were classified to the groups based on their popularity i.e. the most popular word was placed into first group, second popular word into second group, third popular word into third group etc. The aim was to create four groups with words of balanced popularity. With each word from the single group we are able to encode 10 bits of steganogram. To make detection harder each 10 bits sequence is represented by four words (one word randomly chosen from each of four groups). This allowed fulfilling requirement R2. Of course, once the word was randomly selected from the group and assigned to the particular bits sequence it cannot be chosen again to represent different bits sequence.

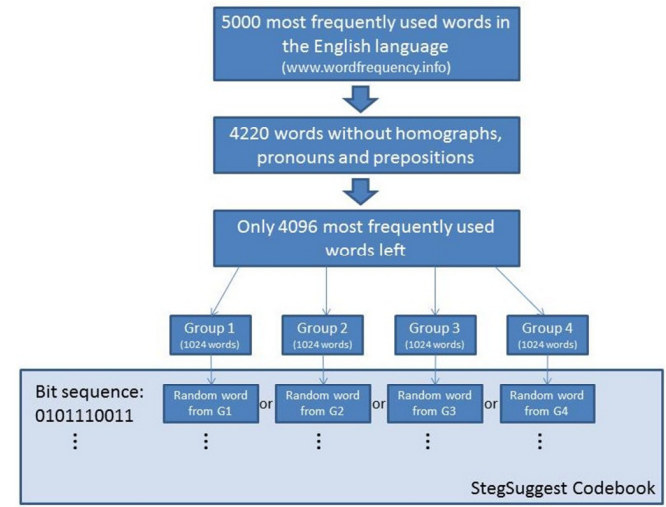

**Fig. 10** StegSuggest codebook design

The designed codebook must be shared between steganograms sender and receiver to properly encode and decode secret data.

**5.3 StegSuggest messages and functioning**

StegSuggest operate based on three types of messages that utilize Google suggestions. They are messages that:
- acknowledge registration of the TCP connection that will be transferring Google suggestions,
- carry steganogram bits,
- close hidden channel.

Each type of the message has its own format. Fig. 3 presented exemplary real Google suggestions for the word 'steganos'. Based on this example, we will show how StegSuggest adds and interprets steg-suggestions. First, notice that steg-suggestions are always added at the end of the single row which contains original Google suggestion. Ten, random StegSuggest words were added to the suggestions from Fig. 3 and are presented in Fig. 11.

```
window.google.ac.h(["steganos",[["steganos\u003Cb\u003E internet anonym
cat\u003C\/b\u003E",0,"0"],["steganos milk",0,"1"],["steganos\u003Cb\u003E security
suite cup\u003C\/b\u003E",0,"2"],["steganos\u003Cb\u003E internet anonym pro 8.0.1
five\u003C\/b\u003E",0,"3"],["steganos\u003Cb\u003E safe search\u003C\/b\u003E",0,"4"],
["steganos\u003Cb\u003E anonym pro find\u003C\/b\u003E",0,"5"],["steganos\u003Cb\u003E
internet anonym pro multilingual v7.0.9 interpret\u003C\/b\u003E",0,"6"],
["steganos\u003Cb\u003E internet anonym download detect\u003C\/b\u003E",0,"7"],
["steganos\u003Cb\u003E internet anonym vpn journal\u003C\/b\u003E",0,"8"],
["steganos\u003Cb\u003E download working\u003C\/b\u003E",0,"9"]],"","","","","",{}])
```

**Fig. 11** Google suggestions with steg-suggestions (in red)

Besides introducing modifications to Google suggestions, StegSuggest also utilises modifications to WS and TS options from TCP header, which will be described in details later in this section.

Now we will describe in what four phases does StegSuggest operate. They are: Phase 1 that requires creation, learning and updating of the Google Suggest servers list, Phase 2 that allows adding Google client-server connection to the list of TCP connections used by StegSuggest, Phase 3 in which steganographic data transfer is taking place and finally Phase 4 in which finishing of the hidden data occurs. The detailed description of each phase is presented below.

**Phase 1:** Creating, learning and updating the Google Suggest servers list.
This phase is necessary for SS and SR to later monitor and influence the TCP connections between Google Suggest client(s) and server(s). It is based on analysing DNS traffic: every time client asks for Google Suggest server, returned IP address is captured and stored until it expires (which is specified in the response DNS message).

**Phase 2:** Adding Google client-server connection to the list of StegSuggest connections.
This is signalling phase in which SS and SR are establishing which Google Suggest connections will be transferring secret data. It requires registering particular TCP connections for steganographic purposes by both: SS and SR. This phase is necessary for the SR to distinguish which Google Suggest responses are carrying hidden data. Proposed TCP connection registration scheme is presented in Fig. 12.

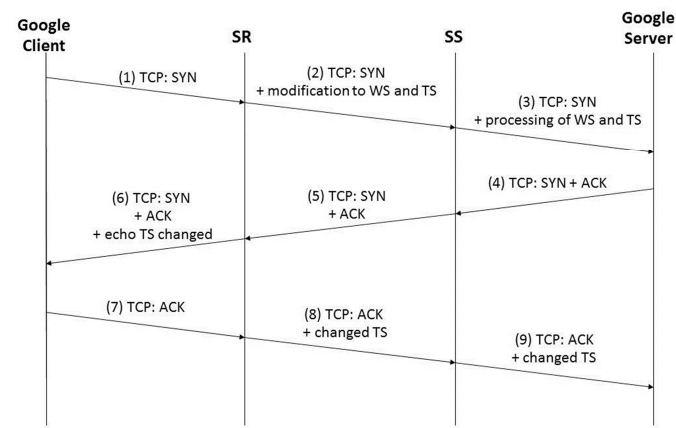

**Fig. 12** Registration request of the TCP connection for steganographic purposes

SR issues TCP connection registration request to the SS each time the TCP connection with one of the Google Suggest server is established. After receiving TCP segment with SYN flag which is directed from client to the Google server (Fig.12, 1), SR modifies TCP options fields: WS and TS. Both these options are used by StegSuggest to signal SS that this connection may be used for steganographic purposes if SS is ready to send secret data. It is then for SS to decide whether it will require new TCP connection for hidden data transfer.

TS option is modified by inserting into the field the value, which will be called a *Steganographic Session Identifier* (*SSI*). SS utilises *SSI* to acknowledge registration of the particular TCP connection. *SSI* which is calcu-lated for the first TCP connection between the SS and SR has a special meaning; it not only permits to register new TCP connection but also creates the hidden channel (which may include one or more registered TCP connections for steganographic purposes). This first *SSI* is called *native SSI*. Its value is stored to be later used while closing the hidden channel. The *SSI* is calculated as

$$SSI = H(ISN \| HCK \| oWS), \qquad (5\text{-}1)$$

where *H* is a cryptographic hash function (e.g. MD-5 or SHA-1). The hash is computed on the following components: *ISN* – initial sequence number of the current TCP segment, *HCK* – a hidden channel key – a number which is a shared secret between SS and SR and *oWS* – original value of WS option that was inserted by Google client (if TS is not present in the TCP segment then *oWS*=15). From the computed hash only 16 bits are chosen to form *SSI*.

The modification of the WS option in the TCP SYN segment is decided based on algorithm presented in Fig 13. Values inserted by StegSuggest into WS option field were chosen based on their frequency occurrence in the real Google traffic (see Sec. 4) – WS that equals 1, 2, 3 and 6 were the most frequently discovered values. Performed real Google traffic analysis also showed that if WS option was not present in the TCP header then also there was no TS option. That is why for all these cases WS=1. In case if both WS and TS are not present in the original TCP segment, then they are added by SR. This requires modification to the TCP *Data Offset* field and shortening of the *Padding*. At the SS side introduced modifications must be reversed.

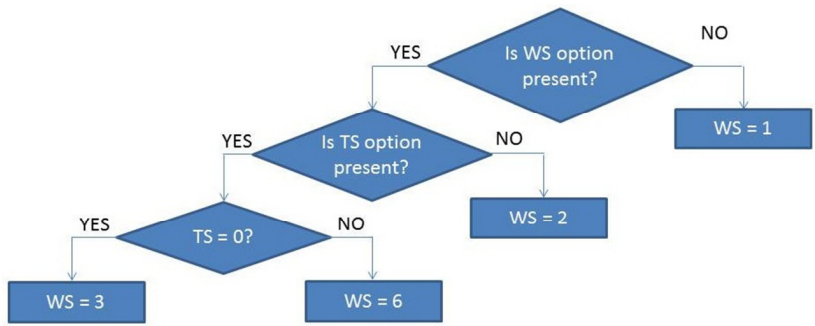

**Fig. 13** WS modification algorithm

After introducing all abovementioned modifications, SR sends the TCP segment to SS (Fig. 12, 2). From this time on, SR must change timestamps in all TCP segments directed from or to Google server. This means that SR saves the original values of the timestamps, and changes them whenever TCP segments are passed. In case that the influenced timestamp value returns to the SR, he/she restores the original one. This ensures that no inconsistency of timestamps is introduced. SR saves four values of the timestamps for segments that are directed to the Goggle server:

- *TSOrigLast* – which stores the original timestamp from the last segment sent to the Google server,
- *TSOrigCur* – which stores the original timestamp from the currently processed segment,
- *TSStegLast* – which stores the timestamp from the last changed segment sent to the Google server,
- *TSStegCur* – which stores the timestamp from the currently processed changed segment.

$$TSStegCur = TSStegLast + (TSOrigCur - TSOrigLast), \qquad (5\text{-}2)$$

*TSStegLast* is inserted by SR into modified segment as an actual timestamp. Google server, in acknowledgment, sends a TCP segment in which *echo timestamp* is then changed at SR to *TSOrigCur* (Fig. 12, 6).

After sending steganographic registration request message, SR waits 5 seconds (which is an arbitrary value) for its confirmation. If registration it is not received (together with first steg-suggestions) then the TCP connection is not used for the steganographic purposes.

After changing TS and WS options SS sends TCP segment into Google server direction (Fig. 12, 3). When the ACK segment is sent in response, it is modified when it reaches SR (Fig. 12, 6) by adjusting TS option value; analogous operation is performed at (Fig. 12, 8) and other similar steps of StegSuggest.

Next, SS based on the presence of the steganogram to send decides whether particular TCP connection will be registered and then used for steganographic purposes. It is achieved by analysing each TCP segment with SYN flag, analysing the WS and TS options in each segment to be more precise. SS verifies whether the TCP segment with SYN flag has proper *SSI* in the TS option. *SSI* is retrieved by performing several calculations based on eq. 5-1 for different WS values. SS discovers the value of the WS and then he/she reverses the algorithm from Fig. 13. For example, if SS finds out that in TCP segment with SYN flag WS=1, then he/she removes WS and TS options and then sends it to the Google Suggest server. If SS discovers that WS=6 then TS is left with-out modification and only the value of the WS is changed to the original one (which is discovered based on the hash inserted by SR into TS option).

The confirmation of the registration of the TCP connection for StegSuggest purposes is performed by SS when the first set of suggestions is sent from the Google Suggest server (Fig. 16). HTTP request which contains the first characters of the search phrase is modified by the SR (by adjusting TS option) and then sent further (Fig. 16, 1) without any changes to Google Suggest server. The server acknowledges the receipt of the request (Fig. 16, 2) and sends the first set of the suggestions (Fig. 16, 3). SS embeds into the first four rows of the suggestions list proper *SSI* (twice) - the same as in the registration of the TCP connection (Fig. 16, 4). The StegSuggest message that confirms registration of the TCP connection is presented in Fig. 14.

```
steganos internet anonym cat
steganos milk
steganos security suite cup
steganos internet anonym pro 8.0.1 five
steganos safe
steganos anonym pro
steganos internet anonym pro multilingual v7.0.9
steganos internet anonym download
steganos internet anonym vpn
steganos download
```

**Fig. 14** Confirmation of the TCP connection registration

In Fig. 14, in the first two rows two steg-suggestions were added (cat and milk): first 16 bits that are encoded with forms *SSI*. The same role has steg-suggestions added in third and fourth rows. The *SSI* is repeated twice to limit the chance of the case when original suggestions are mistaken with *SSI*.

After receiving the first set of suggestions SR verifies the *SSI* by decoding the words from the four rows and comparing with the sent one. If it is correct two *SSIs* are removed and the TCP connection is treated from now on as registered. Next, modified segment is sent to the client (Fig. 16, 5) who confirms the reception of the suggestions with ACK segment (Fig. 16, 6).

For all modified TCP segments for all StegSuggest operations certain fields related to size of the segment must be adjusted accordingly. They are:
- *Total length* field in the IP header,
- *Sequence Number* and *Acknowledgement* in the TCP header,
- HTTP Content-Length in the HTTP header.

**Phase 3:** Steganographic data transfer.
The mechanism of the steganographic data transfer is similar to acknowledgement of the TCP connection registration presented in Phase 2. The only difference is that the SS adds steg-suggestions to the original suggestions sent by Google server.

After receiving Google suggestions SS adds a single additional word to each row. Each of these words carries 10 bits of steganogram (see Sec. 5.2 for StegSuggest codebook creation). It is worth noting that usually there are 10 rows of suggestions generated. But sometimes there are fewer for rare or long search phrases. In such case, some additional fake rows can be inserted to obtain total of 100 bits of steganogram per single set of suggestions. Such fake row will begin with the search phrase and then the code word with steganogram will be added.

When more than single TCP connection (related to Google Suggest service) is utilised by StegSuggest between the SS and SR then steganogram will be inserted by SS into segments with the order of their arrival. However, to ensure proper assembly of the steganogram at the SR side from different segments with steg-suggestions numbering of the parts of the steganogram is required. It is achieved by inserting additional code word in rows 2 and 5. The total of 20 bits available will allow transferring 1 048 576 parts of steganogram. From all values of the sequence numbers only all-zeros value is reserved for closing the hidden channel (see Phase 4). The initial value of the sequence number starts at 1 and when it reaches 1048576 then it is again set to 1. Fig. 15 shows format of the message that carry steganograms. Red steg-suggestions transfer steganograms, green encodes steganogram sequence number.

```
steganos internet anonym cat
steganos milk age
steganos security suite cup
steganos internet anonym pro 8.0.1 five
steganos safe search go
steganos anonym pro find
steganos internet anonym pro multilingual v7.0.9 interpret
steganos internet anonym download detect
steganos internet anonym vpn journal
steganos download working
```

**Fig. 15** StegSuggest message format for steganograms transfer

In the next step, Google suggestions together with steg-suggestions are sent by SS to SR (Fig. 16, 4). After successful reception of the suggestions list, SR first decodes sequence number, removes it and then decodes steganogram (and removes it too). Only original Google suggestions are then sent to the client (Fig. 16, 5).

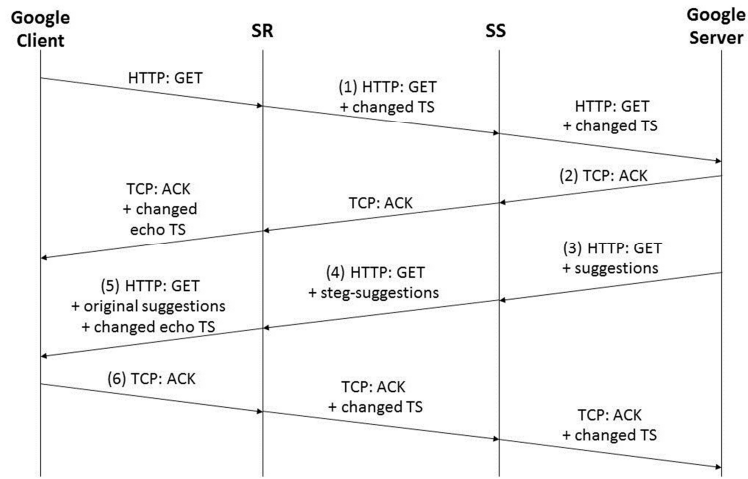

**Fig. 16** Steganogram transfer using StegSuggest

**Phase 4:** Finishing hidden data transfer.
The finish of the hidden data transfer is equivalent with closing the hidden channel. When SS wants to end secret communication he/she issues two signalling messages that are encoded into original Google suggestions. It is two-step process. First, the end of the secret data is signalled by changing the sequence number of the steganogram in the list of suggestions to 0 (as mentioned in Phase 3). Then the message that shuts the covert channel is issued (Fig. 17). When SR discovers such message he/she ceases to insert parts of the steganogram into original suggestions. The overt communication between Google Suggest client and server continues but only timestamps at SR are still modified until particular TCP connection is finished. After that it is deleted from the registered TCP connections list.

```
steganos internet anonym cat
steganos milk
steganos security suite cup
steganos internet anonym pro 8.0.1 five
steganos safe age
steganos anonym pro
steganos internet anonym pro multilingual v7.0.9
steganos internet anonym download
steganos internet anonym vpn
steganos download
```

**Fig. 17** StegSuggest message format for closing hidden channel

Fig. 17 presents format of the message that is devoted to closing the hidden channel. Steg-suggestions in rows: 1, 2 and 3, 4 carry session identifier (in red). In fifth row the steg-suggestion that describes the number of bits sent in the last suggestions list (in green) is transferred (if it is required).

### 5.4 StegSuggest steganographic bandwidth estimation

Based on the analysis of the real Google Suggest traffic that was presented in Section 4, we will now assess potential steganographic bandwidth of the StegSuggest method. Experimental results show that each Google search involved, on average, sending of about 7 suggestion lists. Taking into consideration that each list will carry 100 bits of secret data the average amount of steganogram per Google search is 700 bits. Thus, the more Google traffic is generated by users, the higher steganographic bandwidth. Our experimental results showed that the average user generated single Google search every three hours, which is rather rarely and it gives steganographic bandwidth of about 0.1 bit/s for single user which is not impressive. However, if StegSuggest is used for aggregated Google traffic (and this scenario is considered in this paper – see subsection 5.1) then the steganographic bandwidth increases e.g. for 100 users it is about 10 bit/s. Of course, our captured real network traffic cannot be treated as representative when it comes to Google Suggest service, as there are other network environments were the statistics of this kind traffic will be tremendously different, for example in hot spots, internet cafes or student dormitories. If SS and SR have access to such aggregated traffic then the resulting steganographic bandwidth can be considerably higher.

## 6. Conclusions and future work

In this paper we presented a new steganographic method named StegSuggest which is the first network steganography solution that allows hidden data transfer using Google search. As a hidden data carrier StegSuggest utilises Google Suggest service's suggestions. It was estimated that it allows sending 100 bits of steganogram for every Google suggestions list. When steganogram sender and receiver have access to the aggregated Google Suggest traffic then it achieves decent steganographic bandwidth (about 10 bit/s).

In December 2010 new service was introduced by Google in its search. It is called Google Instant [22] and shows search results while user is still typing the search phrase. It is worth noting that Google Suggest after modifications is capable of utilising also Google Instant and it will presumably achieve higher steganographic bandwidth than for Google Suggest because the volume of data which is sent with every user's keystroke is significantly higher.

Future work will be focused on development of prototype of the StegSuggest to prove the concept. Potential detection methods should also be discussed.

ACKNOWLEDGMENT

This work was partially supported by the Polish Ministry of Science and Higher Education under Grant: N517 071637 and IP2010 025470.